\documentclass[aip,apl,reprint]{revtex4-1}
\usepackage[utf8]{inputenc}
\usepackage{graphicx}
\usepackage{siunitx}				
\usepackage{textcomp}
\usepackage{amsmath}
\begin{document}
\title{Photophoretic trampoline~-- Interaction of single airborne absorbing droplets with light}
\author{Michael Esseling}
\email{michael.esseling@uni-muenster.de}
\author{Patrick Rose}
\author{Christina Alpmann}
\author{Cornelia Denz}
\affiliation{Institut für Angewandte Physik and Center for Nonlinear Science (CeNoS), Westfälische Wilhelms-Universität Münster, Corrensstraße~2/4, 48149~Münster, Germany}
\begin{abstract}
We present the light-induced manipulation of absorbing liquid droplets in air. Ink droplets from a printer cartridge are used to demonstrate that absorbing liquids~-- just like their solid counterparts~-- can interact with regions of high light intensity due to the photophoretic force. It is shown that droplets follow a quasi-ballistic trajectory after bouncing off a high intensity light sheet. We estimate the intensities necessary for this rebound of airborne droplets and change the droplet trajectories through a variation of the manipulating light field.
\end{abstract}
\maketitle
%
%% Introduction
The natural occurrences of emulsions and aerosols and their relevance in many fields like climate research~\cite{Ramanathan2001}, fuel engineering~\cite{Abramzon2006}, or digital microfluidics~\cite{Teh2008,Hung2006} stimulate the modern research on liquids. In many cases, these studies require a controlled handling of liquid droplets~\cite{Krieger2012}. In this context, several manipulation techniques for the confinement of a large variety of droplets have been developed. These include electric~\cite{Kim2011} or magnetic fields~\cite{Okochi2010}, optoelectrowetting~\cite{Park2009}, optical tweezers for transparent droplets~\cite{Hoffmann1993,Burnham2006,McGloin2007}, or acoustic traps~\cite{Davies2000}. However, no method has been demonstrated so far that is able to manipulate airborne absorbing droplets all-optically. While established optical manipulation schemes use gradient forces for the control of transparent particles~\cite{Ashkin1986,Ashkin1987,Baumgartl2008}, this approach is challenging for micro particles or droplets that possess non-negligible absorption\cite{Rubinsztein-Dunlop1998}. Photophoresis, as described by Ehrenhaft~\cite{Ehrenhaft1917} in 1917, has recently been discovered to be a versatile tool for the all-optical control of absorbing matter~\cite{Desyatnikov2009}. Photophoretic forces result from thermal gradients on an object's surface pushing it away from regions of high light intensity~\cite{Beresnev1993}. Based on this idea, several trapping configurations for solids have been proposed, making use of single or multiple hollow beams forming light cages~\cite{Gahagan1996,Desyatnikov2009,Shvedov2010,Zhang2011a,Alpmann2012,Shvedov2012}.

In this work, we introduce the photophoretic manipulation of absorbing airborne droplets. However, the transfer of light-induced thermophoretic forces to the manipulation of absorbing liquids is not straightforward and one has to take into account multiple effects. The built-up of local temperature gradients that are fundamental for these forces is influenced not only by heat diffusion, as in the case of solid particles, hence it may be hampered by convection flows inside the liquid or even premature evaporation of the droplet~\cite{Iwaki2010}. For the generation of droplets, existing experiments frequently use ultrasonic nebulizers that can provide a large number of airborne droplets at the same time~\cite{Walker2011,McGloin2007} having broad statistical distributions in size, velocity, and momentum. In order to prepare a system with defined initial conditions, these aerosols require further sophisticated concepts for the separation and sorting of single droplets~\cite{Horstmann2012}. To produce well-defined droplets, we use a different approach and utilize an inkjet printer head as the droplet generator~\cite{Hoffmann1993}. In our case, the cartridge ejects droplets by heating an internal resistor with very short but intense electrical pulses that rapidly vaporize some of the water-based ink. For each pulse, the increased pressure near the resistor pushes out a high velocity droplet. By this means we are able to generate single droplets of absorbing liquid on demand.

\begin{figure}
\centering
\includegraphics{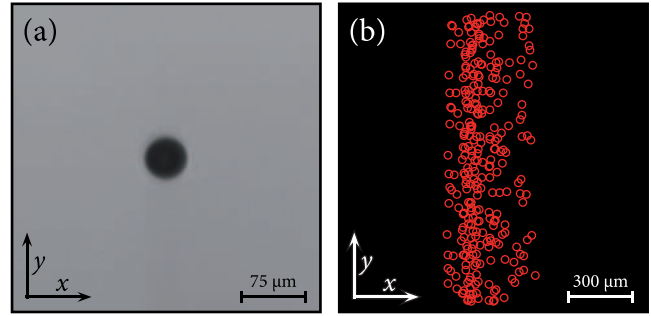}
\caption{(Color online) Characterization of droplet generation. (a)~Representative image of a falling droplet, (b)~tracked centers of 75 falling droplets.}
\label{fig:dropletCharacterization}
\end{figure}

Figure~\ref{fig:dropletCharacterization}(a) shows a representative image of such a single droplet. Analyzing many droplets, we measured a narrow size distribution centered around a mean radius of $r$ = \SI[separate-uncertainty=true]{25 +- 1}{\micro\meter}. It is well-known that in this size regime surface tension dominates over friction which results in the preservation of the spherical shape of a droplet~\cite{Hadamard1911}, as clearly observed in our experiments [cf.\ Fig.~\ref{fig:dropletCharacterization}(a)]. For defined initial conditions with a precise trajectory and a moderate velocity of the droplet, we mount the cartridge to shoot against gravity in the vertical direction and characterize the generated droplets with a high-speed camera. Using the data from our size characterization and a model that includes Stokes friction $F_\mathrm{Stokes} =  6 \pi r \nu u$, where $r$ is the radius, $\nu$ the dynamic viscosity of air at room temperature, and $u$ the velocity, we were able to estimate the trajectory of the emitted droplets. Based on this, we expect the individual droplets to travel a distance of about \SI{4}{\centi\meter} before reaching the turning point. On their way down, they are accelerated until they fall with a terminal velocity of \SI[separate-uncertainty = true]{79.4 +- 6.3}{\milli\meter \per \second}. This value was experimentally verified by tracking multiple droplets each in consecutive frames which yielded a mean terminal velocity of \SI[separate-uncertainty = true]{84 +- 6}{\milli\meter \per \second}.

\begin{figure}
\centering
\includegraphics{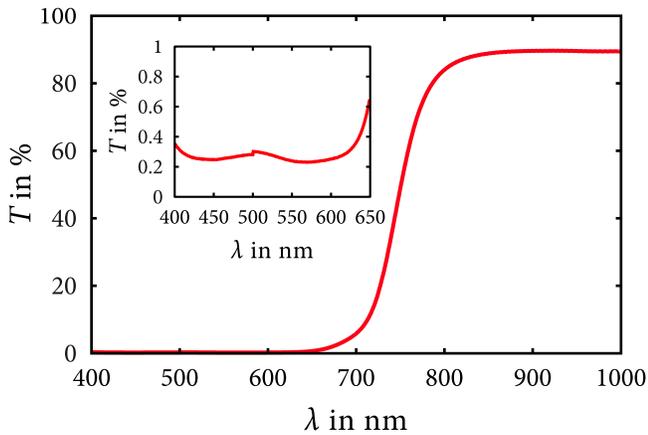}
\caption{(Color online) Fresnel-corrected transmission spectrum through a \SI{60}{\micro\meter} thick layer of the used ink.}
\label{fig:inkSpectrum}
\end{figure}

Figure~\ref{fig:dropletCharacterization}(b) shows the traced centers of 75~falling droplets. This overlay plot underlines the high reproducibility of the individually emitted droplets. The measured trajectories possess little scattering, in particular when considering that particles leave the cartridge at an initial velocity of about \SI{2}{\meter \per \second}. Since photophoresis is an effect that influences absorbing matter only, we checked the used ink for a sufficient absorbance. A thin layer of ink was prepared between two glass slides with a distance of \SI{60}{\micro\meter} and the corresponding transmission spectrum from \SI{400}{\nano\meter} to \SI{1000}{\nano\meter} was measured using a commercially available spectrometer. Figure~\ref{fig:inkSpectrum} shows the obtained spectrum after correction for the Fresnel reflections at the air-glass- and glass-ink-interfaces. We clearly see a high absorption in the whole visible range. In particular, we estimate an absorption coefficient of $\alpha = \SI[separate-uncertainty = true]{0.09 +- 0.01}{\per\micro\meter}$ for a wavelength of \SI{532}{\nano\meter}.

\begin{figure}
\centering
\includegraphics{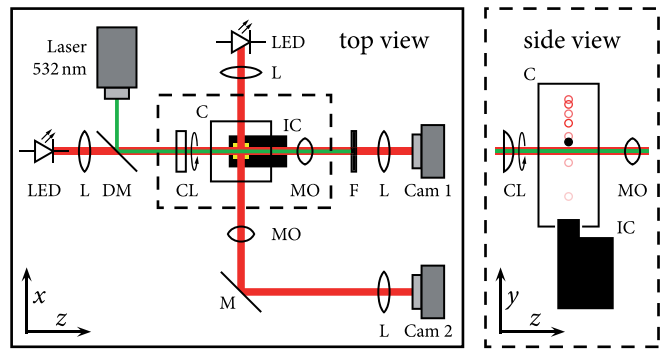}
\caption{(Color online) Schematic setup for the photophoretic manipulation of droplets. (Left)~Top view with two perpendicular microscopes centered around the inkjet cartridge, (right) side view of cartridge and shielding glass cuvette allowing for a reproducible droplet trajectory illustrated with a streak series of traced droplet centers. C:~cuvette, CL:~cylindrical lens, DM:~dichroic mirror, F:~longpass filter, IC:~inkjet cartridge, L:~lens, LED:~light-emitting diode, M:~mirror, MO:~microscope objective.}
\label{fig:setup}
\end{figure}

\begin{figure}
\centering
\includegraphics{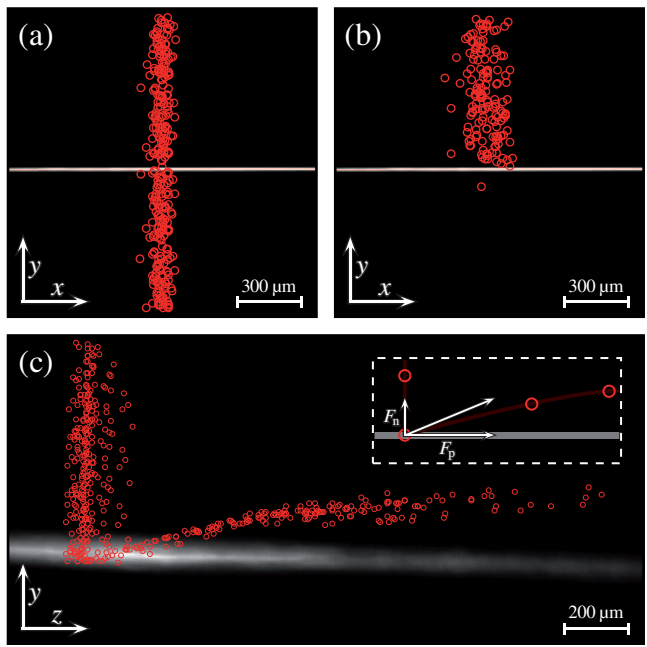}
\caption{(Color online) Photophoretic manipulation of absorbing droplets. (a)~Tracked centers of 75~droplets falling through a light sheet with a peak intensity of $I_\mathrm{peak} \approx \SI{57}{\micro \watt \per \square\micro\meter}$, (b)~tracked centers of 75~droplets repelled by a light sheet of $I_\mathrm{peak} \approx \SI{115}{\micro \watt \per \square\micro\meter}$ (cf.\ also movie1.mpg), (c)~side view showing the axial movement of repelled droplets (cf.\ also movie2.mpg), the light sheet is visualized with a normalized image of the averaged scattered light from an aerosol of solid particles, the inset schematically illustrates the two acting force components (enhanced online).}
\label{fig:axialMomentum}
\end{figure}

For the investigation of the photophoretic manipulation of absorbing droplets, the droplet generator was incorporated into the laser setup shown schematically in Figure~\ref{fig:setup}. A frequency-doubled Nd:YAG laser ($\lambda = \SI{532}{\nano \meter}$) was focused above the inkjet cartridge by a cylindrical lens with a focal length of $f = \SI{60}{\milli\meter}$. For the observation, we used two microscopes that imaged the droplets in the transverse $xy$- and the perpendicular $yz$-plane, respectively. Images were captured with a frame rate of \SI{240}{fps} using a hardware-triggered high-speed camera synchronized to the droplet emission signal. The microscope illumination was provided by light emitting diodes, while the laser light was blocked with a longpass filter. To ensure the high reproducibility of droplet trajectories, the observation volume was encapsulated by a glass cuvette blocking air flows in the laboratory. Additionally, droplets were shot with a temporal separation of \SI{20}{\second} so that air fluctuations induced by one droplet have enough time to settle before the next one is emitted.
Since the cylindrical lens focuses the laser light in the vertical direction ($y$-direction) only, the obtained light field resembles a plane of light which we call light sheet. It provides a strong intensity gradient in the direction of droplet movement, which is expected to result in an effective photophoretic interaction with the falling droplets. To estimate the laser intensity necessary for the manipulation of the absorbing droplets, we investigate the interaction of falling droplets with light sheets of increasing total power. Figure~\ref{fig:axialMomentum} summarizes the corresponding measurements. Each subfigure combines a microscopic image of the light field with streak series of 75~falling droplets illustrated with red circles indicating the found droplet center positions.

During the measurement, we increased the power in several steps. In the case of no laser illumination, all droplets fell right through the field of view [cf.\ Fig.~\ref{fig:dropletCharacterization}(b)] and Figure~\ref{fig:axialMomentum}(a) shows that even for a total power of \SI{900}{\milli\watt} no significant influence on the falling droplets could be observed. For higher intensities though, we found a photophoretic interaction between light sheet and liquid droplet. In contrast to solid particles, where solely the heat transfer to the surrounding medium is relevant, liquids require a proper treatment of evaporation effects as well~\cite{Iwaki2010}, and indeed we observed little uprising wads of vapor after a droplet impact. As depicted in Figure~\ref{fig:axialMomentum}(b), for a total power of more than \SI{1.8}{\watt}, droplets do not fall through the light sheet any more. Instead, we observed in the experiment that above a certain threshold power they are repelled and seem to be accelerated out of the microscope's focal region.

Our interpretation was corroborated by the side view of the experiment illustrated in Figure~\ref{fig:axialMomentum}(c). This perspective revealed a strong acceleration that is separable into two components, one normal to the light sheet and one in the direction of light propagation [see schematic inset in Fig.~\ref{fig:axialMomentum}(c)]. Due to the short interaction time between droplet and light field as well as the droplets' high axial velocity, it was not possible to resolve the interaction with our camera equipment, capable of aquiring 240 fps. Typically, we obtained only one or two images of each droplet after bouncing off the light sheet even though we used a smaller magnification to enlarge the field of view. Nevertheless, by overlaying all droplet positions, we visualized the parabolic movement of the droplets. Fitting this trajectory with the Stokes friction model, we found an initial normal velocity of $u_{\mathrm{n}} = \SI{65}{\milli\meter \per \second}$ and a velocity of $u_{\mathrm{p}} = \SI{320}{\milli\meter\per\second}$ in the direction of light propagation. Based on these values, we used Newton's Law to estimate the occurring forces to be at least $\textbf{F}_{\mathrm{n}} = \SI{4.8}{\nano \newton}$ and $\textbf{F}_{\mathrm{p}} = \SI{10.4}{\nano \newton}$, respectively, if we assume an underlying interaction time of not more than \SI{2}{\milli \second}. It should be noted that these values are only a very rough estimate of the lower limit for the occuring forces, based on the simplified assumption that during a single interaction event, the photophoretic force is constant. Nevertheless, it can be seen that these rough estimates already yield a photophoretic force two orders of magnitude larger than in typical optical tweezers~\cite{Malagnino2002}. As mentioned above, the photophoretic effect leads to a force pointing away from regions of higher light intensity. Therefore, the force component normal to the light sheet is directly determined by the shape and orientation of the light sheet whereas the intensity gradient in direction of propagation is due to the geometrical shadow of the droplet entering the illuminated volume. The same effect has been used previously for the transport of solid particles in a vortex beam~\cite{Shvedov2010}. To quantify the peak intensity in the focal plane, we also measured the size of the collimated Gaussian beam in front of the cylindrical lens. Using Gaussian transformation optics along one direction only, one can calculate the form and extent of the light sheet in the focal plane which can be descibed as a Gaussian beam $I(x,y) = I_\mathrm{peak} \exp(-x^2/\omega_x^2)\exp(-y^2/\omega_y^2)$ with different beam waists $\omega_x =$ \SI{1.23d-3}{\meter} and $\omega_y =$ \SI{8.29d-6}{\meter} in horizontal and vertical direction, respectively. Together with the information about the incident power of \SI{1.8}{\watt}, a peak intensity of $I_\mathrm{peak} \approx$ \SI{115}{\micro \watt \per \square\micro\meter} in the focal plane could be derived. Since the light propagation was perpendicular to the direction of observation in Figure~\ref{fig:axialMomentum}(c), a direct image of the laser field could not be obtained. Instead, the light sheet image was visualized by averaging and then normalizing a video sequence of scattering particles in a second, closed glass cuvette. A slight displacement exists, however, between the so-estimated intensity distribution and the measured turning points of droplets. This is probably due to the fact that between these experiments, the glass cuvette had to be changed to the one containing the scattering particles, causing a beam displacement.

\begin{figure}
\centering
\includegraphics{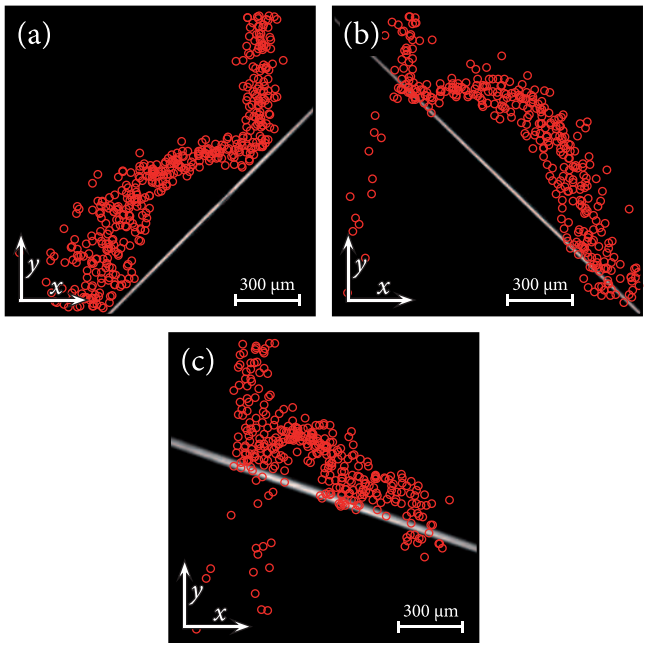}
\caption{(Color online) Photophoretic trampoline with a total intensity of \SI{1.8}{\watt}. (a)~Transverse bouncing of airborne droplets towards the left off a light sheet with inclination angle $\beta \approx \SI{45}{\degree}$, (b)~manipulation towards the right with a light sheet inclination of $\beta \approx \SI{-45}{\degree}$, (c)~multiple droplet bouncing off a $\beta \approx \SI{-20}{\degree}$ light sheet (cf.\ also movie3.mpg) (enhanced online).}
\label{fig:transverseMomentum}
\end{figure}

As the direction of the total photophoretic force depends on the shape of the manipulating intensity distribution, we were able to alter the droplet trajectories over a wide range by changing the light field. The cylindrical lens was installed in a rotation mount which allowed a continuous inclination of the angle $\beta$ between the light sheet and the horizontal plane. Since this reorientation allowed us to induce an additional force component in the $x$-direction, we analyzed the droplet motion in the $x$-$y$-plane again (cf.\ Fig.~\ref{fig:transverseMomentum}). Figure~\ref{fig:transverseMomentum}(a) shows a light field with an inclination of $\beta \approx \SI{45}{\degree}$. The falling droplets now bounced not only in the upright direction but got a horizontal momentum as well leading to a parabolic movement in the left direction as can clearly be seen by the tracked droplet centers. Consequently, an inclination of $\beta \approx \SI{-45}{\degree}$ will lead to a corresponding transport to the right which actually can be observed in Figure~\ref{fig:transverseMomentum}(b). Provided that the light field can be changed rapidly, such an arrangement allows for a very efficient sorting concept for liquid droplets. Furthermore, this experiment demonstrates again the intensity dependence of the photophoretic effect. If droplets miss the region of sufficient light intensity, they might fall through the sheet and gain a transverse momentum to the opposite direction [cf.\ Fig.~\ref{fig:transverseMomentum}(b)]. Moreover, the number of interactions is not limited to just one event. Figure~\ref{fig:transverseMomentum}(c) highlights the occurrence of multiple droplet impacts in the field of view. For an inclination angle of $\beta = \SI{-20}{\degree}$, we observe up to three accelerations of the droplet in the direction normal to the light sheet~-- leading to a continuous bouncing on the photophoretic trampoline.

%% Conclusion
In conclusion, we demonstrated the photophoretic manipulation of absorbing liquids. In order to realize single airborne droplets with highly reproducible initial conditions, we utilized a computer-controlled inkjet cartridge as droplet generator, a concept that is feasible for many substances~\cite{Hoffmann1993}. For highly absorbing ink droplets, we estimated the intensity necessary for a photophoretic manipulation to be in the range of about \SI{100}{\micro\watt\per\square\micro\meter}. However, this value is expected to be highly dependent on both the absorption of the liquid and the interaction time of droplet and light field. An analysis of the interaction between absorbing micro-droplets and a light sheet revealed two relevant components of the photophoretic force, one acting in the direction of light propagation, the other one in the direction normal to the light plane. Since the latter component depends on the shape and orientation of the light field, we were able to control the droplet trajectories. By changing the inclination of the focused light sheet, we achieved continuous bounces of the droplet on the photophoretic trampoline. Adopting recent developments in tailored light field generation~\cite{Rose2012}, this will lead to an enormous flexibility concerning complex manipulation and trapping geometries. For example the integration of a spatial light modulator might enable the implementation of hollow light fields like holographic optical bottle beams~\cite{Alpmann2012} for microfluidic handling. We strongly believe that the demonstrated photophoretic droplet manipulation will inspire many interdisciplinary research projects and applications in fields like optofluidics, microchemistry, or biomedical science.

\vspace{2mm}
The authors gratefully acknowledge financial support from the Deutsche For\-schungs\-ge\-mein\-schaft in the frame of the Chinese-German transregional research project TRR61.

\end{document}